\documentclass[letterpaper, 10 pt, conference]{ieeeconf}

\IEEEoverridecommandlockouts                              
\usepackage{graphics} % for pdf, bitmapped graphics files
\usepackage{epsfig,epstopdf} % for postscript graphics files
\usepackage{times} % assumes new font selection scheme installed
\usepackage{amsmath,amsfonts,bm} % assumes amsmath package installed
\usepackage{moresize} % ssmall for between tiny and scriptsize
\usepackage{amssymb}  % assumes amsmath package installed
\usepackage[caption=false,font=footnotesize]{subfig}
\usepackage{cite}
\usepackage{flushend}
\usepackage{latexsym} % add box at the end of the line using \hfill
\usepackage{color}
\usepackage{algpseudocode, algorithm}

\DeclareMathAlphabet{\mathcal}{OMS}{cmsy}{m}{n}

\makeatletter
	\renewcommand*\env@matrix[1][c]{\hskip -\arraycolsep
	\let\@ifnextchar\new@ifnextchar
 	\array{*\c@MaxMatrixCols #1}}
\makeatother

\newtheorem{example}{Example}
\newtheorem{theorem}{Theorem}
\newtheorem{lemma}{Lemma}
\newtheorem{prop}{Proposition}

\title{\huge
A Global Identifiability Condition for Consensus Networks with Tree Graphs }

\author{Seyedbehzad Nabavi$^{1}$, Aranya Chakrabortty$^{1}$, and Pramod
 P. Khargonekar$^{2}$
\thanks{This work is supported by National Science Foundation under Grant No. ECCS1062811}
\thanks{$^{1}$S. Nabavi and A. Chakrabortty are with Department of Electrical and Computer Engineering, North Carolina State University, Raleigh, NC, Emails:
        {\tt\small snabavi@ncsu.edu} and {\tt\small achakra2@ncsu.edu}}%
\thanks{$^{2}$P. P. Khargonekar is with Department of Electrical and Computer Engineering, University of Florida, Gainesville, FL, Email:
        {\tt\small ppk@ufl.edu}}%
}

\begin{document}
\maketitle
\thispagestyle{empty}
\pagestyle{empty}

%%%%%%%%%%%%%%%%%%%%%%%%%%%%%%%%%%%%%%%%%%%%%%%%%%%%%%%%%%%%%%%%%%%%%%%%%%%%%%%%

%%%%%%%%%%%%%%%%%%%%%%%%%%%%%%%%%%%%%%%%%%%%%%%%%%%%%%%%%%%%%%%%%%%%%%%%%%%%%%%%%
\begin{abstract}
%\boldmath
In this paper we present a sufficient condition that guarantees identifiability of linear network dynamic systems exhibiting continuous-time weighted consensus protocols with acyclic structure. Each edge of the underlying network graph $\mathcal G$ of the system is defined by a constant parameter, referred to as the \emph{weight} of the edge, while each node is defined by a scalar state whose dynamics evolve as the weighted linear combination of its difference with the states of its neighboring nodes. Following the classical definitions of \emph{identifiability and indistinguishability}, we first derive a condition that ensure the identifiability of the edge weights of $\mathcal G$ in terms of the associated transfer function. Using this characterization, we  propose a sensor placement algorithm that guarantees identifiability of the edge weights. We describe our results using several illustrative examples.
\end{abstract}
\section{Introduction}
\label{Sec_1}
%Over the past decade, there has been significant progress in research on multi-agent networked dynamic systems (NDS) including modeling, analysis, estimation and control, see for example, \cite{reza,ali,murray,chap,nabii,shahram,sandberg} and references cited there. There have also been  a variety of applications in robotics, sensor networks, aerospace systems, communication networks, power systems, and biological networks, see for example, \cite{magnus,bullo,derek,aranya,behzad}. Seminal results on the design of feedback control systems for NDS have been developed using ideas of cooperative control \cite{ren}, passivity \cite{murat}, leader-follower adaptation \cite{he}, and observer-based control \cite{bullo2,chen}, with associated results on controllability and observability of NDS in terms of their graph-theoretic properties \cite{mes1,mes2, nabi}. 

In order to design and analyze monitoring and control algorithms for a networked dynamic system (NDS) using model based approaches, system identifiability is an important question, i.e., whether the dynamic model of the network can be identified uniquely using  available input-output data. This is particularly true for safety-critical networks such as aerospace systems \cite{bullo} and power systems \cite{aranya}, where model parameters change  due to changes in operating conditions, loads, traffic congestion, and network topologies. Consequently, the network model needs to be identified on suitable time scales so that control decisions can be made based on the relevant model. For example, as shown in \cite{aranyatsg}, operators of large power systems typically prefer updating their system models every ten to twenty minutes so that they can predict and control the oscillations in the power-flows with highest accuracy.
% For this, they must keep continuous track of the sensors and their locations in the network graph as losing a bus or a transmission line may result in the loss of identifiability of the desired model. Traditionally speaking,  system operators have often bypassed these modeling problems by resorting to state estimation that only require geometric and dynamic observability of the network \cite{abur}. %both of which, as shown in this paper, are mere subcases of identifiability. 
%With the increasing number of sensors in all such networks, the problem of selecting a sufficient set of sensors that 

Clearly, model identification depends on input-output data which in turn depends on the placement of sensors in the network. A key question therefore, is - on which nodes in the network should one place sensors so that the resulting measurements can guarantee  identifiability of the network model?
%The problem of placing sensors at the right nodes in such networks to guarantee unique identifiability of the network model is, therefore, becoming imperative. %infusion of sensing and communications technologies,  development of high-fidelity dynamic models of NDS is becoming more feasible with the potential for improved performance.
\textcolor{black}{In this paper we answer this question for linear consensus networks defined over acyclic or tree-structured graphs. Such graphs are commonly encountered in  power systems \cite{sce} and social networks \cite{social,sullivant}.} Each edge of the network graph $\mathcal G$ is defined by a constant positive parameter, referred to as the {\it weight} of the edge, while each node is defined by a scalar state whose dynamics evolve as the weighted linear combination of its difference with the states of its neighboring nodes. Our goal is to find a set of outputs from which the edge weights of $\mathcal{G}$ can be identified uniquely. Using the classical notion of identifiability, we frame the problem in terms of relating these edge weights with the Markov parameters of the model, or equivalently, with its transfer function.
%of identifiability of the edge weights in terms of the Markov parameters of the model, or equivalently its transfer function. 
Thereafter, we propose a sensor placement algorithm that guarantees global identifiability of the edge weights from the resulting input-output measurements. We also derive an expression for the number of sensors needed to uniquely identify a complete set of edge weights.

\begin{figure}[!b]
\centering 
\subfloat[Network 1]{\includegraphics[scale=.71]{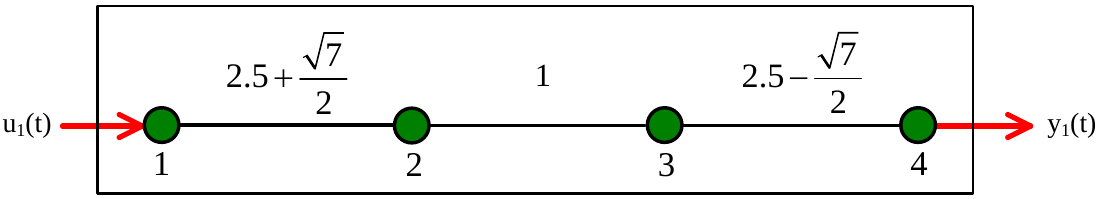}\label{fig01}}\\
\subfloat[Network 2]{\includegraphics[scale=.71]{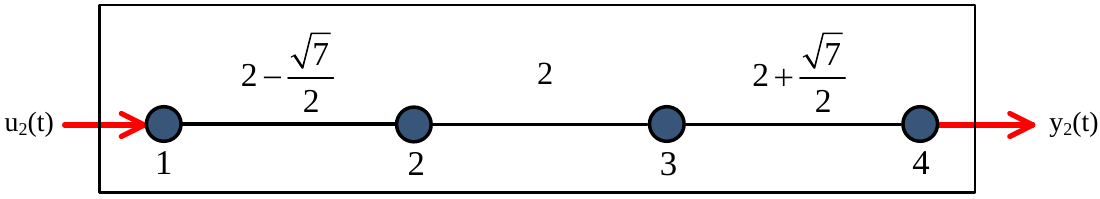}\label{fig02}}
\caption{Two networks with the same input-output transfer function as shown in (\ref{0})}
\label{fig0}
\end{figure}

We wish to emphasize that our objective is not to derive  network identification algorithms such as those in \cite{behzad,shahram,nabi}. Rather, our goal is to find a sufficient set of nodes in a consensus graph where sensors should be placed so that the  measurements available from these sensors may allow one to identify the edge weights of a graph uniquely. Analysis of identifiabilty is the first step in identification process before selection of a specific identification algorithm. To illustrate the fact that this is non-trivial, consider two line graphs shown in Figs. \ref{fig01} and \ref{fig02}.  The graphs have different edge weights but their input-output transfer functions are both equal to:

\vspace{-1em}
\begin{small}
\begin{align}
\frac{Y_1(s)}{U_1(s)}=\frac{Y_2(s)}{U_2(s)}=\frac{4.5}{s^4+12s^3+33s^2+18s}.
\label{0}
\end{align}
\end{small}%
Therefore, regardless of the specific identification algorithm, it is impossible to distinguish these two graphs from each other from the input-output data. Our objective is to develop a sensor placement strategy that guarantees this distinguishability globally over all admissible parameter values defining the edge weights of tree graphs such as those in Networks 1 and 2.

% for any given graph with tree structure. Our examples show that for several tree graphs, the algorithm reduces to both necessary and sufficient conditions for identifiability. Even otherwise, only a small number of additional sensors are needed to guarantee necessity.

Identifiability of linear-time invariant (LTI) systems has been studied for the last several decades \cite{bellman,Ljung2,cor,Grewel,vander,sullivant} and many standard textbooks on the topic are available. 
%%a long-standing literature starting, for example, with the seminal paper by Bellman and \AA str\"{o}m \cite{bellman} which introduced the concept of structural identifiability, and showed that if a LTI system has full state observability and controllability, then it is identifiable. 
%Results on global parameter identifiability were presented by Ljung and Glad in \cite{Ljung2} and  Glover and  co-authors  \cite{cor,Grewel} with an additional rank-test to check local identifiability using {\it Markov} parameters in \cite{Grewel}. Other necessary conditions for identifiability of structured linear systems using Markov parameters were presented in \cite{vander,sullivant}. 
However, not many of these results have been translated to identifiablity of a network, and even less so from a graph-theoretic point of view. This is because  the conventional tests of identifiability, which are mostly formulated as rank tests of Jacobian matrices, quickly become intractable when one attempts to interpret them in terms of the properties of a graph. Some important research efforts have been made in the recent work of \cite{aghdam}, which provide necessary and sufficient conditions for identifiability of consensus networks with unweighted digraphs. Extensions of these necessary conditions have been presented in \cite{salapaka} for graphs with random processes. Relationships between the transfer function of an unweighted graph and its structural properties (such as  the number of the spanning trees) have been presented in \cite{nabii}. Another interesting result is presented in  \cite{roy}, where the objective is to \emph{detect} the loss of an edge in a graph using statistical estimation methods such as maximum a posteriori estimation. In contrast to these results, in this paper we present a sufficient condition on identifiability of weighted graphs from a completely geometric point of view. 
%These conditions, however, do not hold for weighted graphs in general. \textcolor{black}{In a recent work \cite{roy}, the identifiability of link-failures  using noisy measurements is studied by formulating a maximum a-posteriori problem. In \cite{nabii} some relationship between the spectral and structural properties of a SDN defined over unweighted graphs is discussed without investigating the unique identifiability of the underlying network. In contrast to these results, our work in this paper presents a sufficient condition on sensor placement to ensure identifiability of the weights of a network from input-output data. Our analysis takes a geometric point of view by interpreting Markov parameters in light of results from algebraic graph theory \cite{godsil}. 
Specifically, we derive an explicit functional relationship between the Markov parameters and the graph Laplacian matrix, and show how the Laplacian structure relates to the mapping between the Markov parameters and the edge weights.  

The remainder of the paper is organized as follows: Section~\ref{sec2} introduces some preliminaries and notations from graph theory used throughout the paper. Sections \ref{S3} and \ref{S4} provide the main results including the proposed sensor placement algorithm with supporting lemmas. Section \ref{sec_ex} provides some examples. Section \ref{sec_star} provides a note on star graphs.  Section \ref{con} concludes the paper. 
\section{Preliminaries}
\label{sec2}
A \emph{weighted undirected graph} $\mathcal{G}$ consists of three sets: a vertex (node) set $\mathcal{V}$, an edge set $\mathcal{E} \subseteq  \mathcal{V} \times \mathcal{V}$, and an edge-weight set denoted by $W$ which assigns a real positive  number $w_{uv}=w_{vu}$ to all $u \sim v$, where $u \sim v$ is a notational convention for $uv \in \mathcal{E}$. $\mathcal{N}_v$ denotes the set of all nodes neighboring node $v$. The edge joining two nodes $u$ and $v$ is said to be \emph{incident} on them, and is denoted as $uv$. The weight of this edge is denoted as $w_{u,v}$. %The \emph{degree} of the node $v$ denoted by $d_v$ is the summation of the weights of all edges that are incident to $v$. 
A {\it walk} from $u$ to $v$ is a sequence of vertices starting from $u$ and ending to $v$ in which any two consecutive vertices are adjacent. A {\it path} from $u$ to $v$ is a walk in which no vertex is repeated. The \emph{weight} of the path $P$ denoted by $\mathcal{W}(P)$ is equal to the product of its edge weights. A {\it cycle} in a graph is a walk with no repeated vertex other than the beginning and the end. A graph $\mathcal{G}$ is called {\it connected} if there exists a path between any two arbitrary nodes. The {\it distance} between nodes $u$ and $v$ denoted by $d(u,v)$ is the number of the edges of the minimum path connecting $u$ and $v$. A \emph{tree} $\mathcal{T}$ is a connected graph with no cycle. A \emph{rooted graph} is a graph with a special node labeled as {\it root} and all other nodes are ordered with respect to the root. A {\it leaf} in a tree is a vertex which has only one neighbor. %A \emph{cycle graph} $\mathcal{C}_k$ is a graph that consists of only one cycle passing through all $k$ nodes.   

The {\it weighted Laplacian matrix} of graph $\mathcal{G}$ denoted by $L$ is defined as

\vspace{-1em}
\begin{small}
\begin{align}
[L]_{i,j}=\left\{\begin{matrix}
-w_{i,j} ~& i \sim j \\ 
\sum_{k \in \mathcal{N}_i} w_{i,k} & i=j\\
0 & \mathrm{otherwise}
\end{matrix}\right. .
\label{1}
\end{align} 
\end{small}%
where $[.]_{i,j}$ denote the $(i,j)^{th}$ element of a matrix. It can be verified that the matrix $L$ is symmetric for undirected graphs~\cite{godsil}. 

\section{Problem Formulation}
\label{S3}
\noindent Consider a single-input linear consensus network model of the form

\vspace{-1em}
\begin{small}
\begin{align}
\dot{\bm{x}}(t)=\mathcal{L}(W)\bm{x}(t)+Bu(t),~\mathbf{y}(t)=C\bm{x}(t),~\bm{x}(0)=0, \label{eq1}
\end{align}
\end{small}%
where $\bm{x}(t) \in \mathbb{R}^{n}$ is the state, $u(t) \in \mathbb{R}$ is the input, $\mathbf{y}(t) \in \mathbb{R}^{h}$ is the output, the state matrix $\mathcal{L}(W)=-L \in \mathbb{R}^{n\times n}$, where $L=L^T$ is the symmetric graph Laplacian of a network with a given topology $\mathcal{G}$ but unknown edge-weights $W=\{w_{q,v}~|~q\sim v\}$. Let the input node be labeled arbitrarily as node 1, i.e., $B=\mathbf{e}_1 \in \mathbb{R}^{n \times 1}$, where $\mathbf{e}_j$ denote the indicator unit vector whose $j^{th}$ element is 1 and the remaining elements are 0, and $C \in \mathbb{R}^{h \times n}$, where the rows of $C$ are indicator vectors implying that the outputs are a set of specific states. Our objective is to find the matrix $C$ such that, for any given $\mathcal{G}$, we can uniquely identify the parameter vector $W$ from the available state measurements. We develop a node selection algorithm such that if sensors are placed at these selected nodes, then $W$ is uniquely identifiable. We first state the basic definition of identifiability in terms of parameter indistinguishability introduced in \cite{Grewel} as the starting step for our objective. 

\noindent {\bf Definition 1:} \emph{Indistinguishability and Identifiability} \cite{Grewel}-- Consider two consensus models (\ref{eq1}) for two parameter vectors $W$ and $W'$, let $u$ and $u'$ denote the inputs, and let $y$ and $y'$ denote the outputs of these two systems. These two paraemeter sets are called \emph{indistinguishable} if for all $u=u'$, $y=y'$. If $W$ and $W'$ are not indistinguishable, they are simply referred to as \emph{distinguishable}. A parameter vector $W$ is said to be \emph{globally identifiable} if for all $W' \neq W$, $W$ and $W'$ are distinguishable. 

A necessary and sufficient condition for indistinguishability is as follows \cite{Grewel}: the parameter vectors $W$ and $W'$ are indistinguishable if and only if 
\begin{align}
C\mathcal{L}^{\ell}(W)B=C\mathcal{L}^{\ell}(W')B,~~\ell \geq 0.
\label{eq_w}
\end{align} 
A parameter vector $W$ is said to be \emph{globally identifiable} if (\ref{eq_w}) implies $W=W'$ \cite{Grewel}. 

\textcolor{black}{
{\bf Remark: } Based on Definition 1, a parameter vector $W$ is identifiable if the mapping from $W$ to the  transfer function from $u$ to $\mathbf{y}$ is one-to-one. Consequently, identifiability depends only on the controllable/observable subsystem.}

In the following section we derive an algorithm to design the output matrix $C$ in (\ref{eq1}) that guarantees global identifiability of $W$ following from Definition 1.

\section{A Sensor Placement Algorithm Guaranteeing Identifiability of Edge Weights}
\label{S4}
\subsection{Main Result}
We start by assuming that $\mathcal{G}$ in (\ref{eq1}) is a rooted tree graph $\mathcal{T}$ with the root node labeled as 1. This root node is also assumed to be the node where the input $u(t)$ enters. Let $p$ denote the length of the longest path from any node of $\mathcal{T}$ to the root. One can then define generations $S_0,S_1,\ldots,S_p$ as subsets of $\mathcal{V}(\mathcal{T})$ such that 
\begin{align}
S_i=\{v \in \mathcal{V}(\mathcal{T}):~ d(v,1)=i\}.
\label{eq_gen}
\end{align}
If $v_i\in S_k$ and $v_j \in S_{k+1}$ and $v_j$ is a neighbor of $v_i$, then $v_i$ is referred to as \emph{the parent} of $v_j$, and $v_j$ a \emph{child} of $v_i$. If multiple nodes have a common parent, then they are referred to as \emph{siblings}. It can be shown that $S_j$ can be partitioned into $|S_{j-1}|$ sets of nodes that are siblings, where $|.|$ denotes the number of elements of a set. $S_j^k$ denotes the set of nodes in $S_j$ that are the children of the node $k \in S_{j-1}$. For example, for the $\mathcal{T}$ shown in Fig. \ref{fig_1b}, $S_0=\{1\},~~S_1=\{2\},~~S_2=\{3,5\},~~S_3=\{4,6\}$. Also, node $1$ is the parent of node $2$, node $2$ is the only child of node $1$, and nodes $4$ and $6$ are siblings. 

Using these definitions, we next state the following two lemmas to construct the proposed sensor placement algorithm. For all the results from this point onward, we will use $[\mathcal{L}]_{i,j}=[\mathcal{L}]_{j,i}=w_{i,j}=w_{j,i}$ interchangeably following from (\ref{1}) and the definition of $\mathcal{L}$. Also, we will drop the argument of $\mathcal{L}(W)$ and simply denote it as $\mathcal{L}$.
\begin{lemma}  
\label{lemma2}
If $\mathcal{G}=\mathcal{T}$ in (\ref{eq1}), where $\mathcal{T}$ is a rooted tree, then the following relationship holds:
\begin{align*}
[\mathcal{L}^k]_{i,1}=\left\{\begin{matrix}
0 & 0 \leq k \leq d(i,1)-1\\ \mathcal{W}(P_{i,1}) & k=d(i,1) 
\end{matrix}\right.
\end{align*} 
where $P_{i,1}$ is the unique path of length $d(i,1)$ connecting nodes $i$ and $1$.
\end{lemma}
\proof The proof follows from the induction on $k$.
\begin{itemize}
\item $k=1$: If $d(i,1)\geq 2$, that is, nodes $i$ and $1$ are not neighbors, then  $[\mathcal{L}^1]_{i,1}=0$. If $d(i,1)=1$, that is, nodes $i$ and $1$ are neighbors, then $[\mathcal{L}^1]_{i,1}=w_{i,1}=\mathcal{W}(P_{i,1})$.  

\item $k=n>1$ where $n$ is an integer: By induction we assume that $[\mathcal{L}^n]_{i,1}=0$ for all nodes $i$ where $d(i,1) \geq n+1$. Also, we assume $[\mathcal{L}^n]_{l,1}=\mathcal{W}(P_{l,1})$ for all nodes $l$ where $d(l,1)=n$.

\item $k=n+1$: We next consider an arbitrary node $j$ where $d(j,1) \geq n+2$. Following the definition of matrix product $\mathcal{L}^k=\mathcal{L}\mathcal{L}^{k-1}$, one can write the following relationship for any arbitrary node $v \in \mathcal{V}$
\begin{align}
[\mathcal{L}^{k}]_{v,1}=[\mathcal{L}]_{v,v}[\mathcal{L}^{k-1}]_{v,1}+\sum_{\ell \in \mathcal{N}_v}[\mathcal{L}]_{v,\ell}[\mathcal{L}^{k-1}]_{\ell,1}.
\label{eq_im}
\end{align}
Therefore, using (\ref{eq_im}), we can write the following relation for node $j$
\begin{align}
[\mathcal{L}^{n+1}]_{j,1}=&[\mathcal{L}]_{j,q}[\mathcal{L}^{n}]_{q,1}+[\mathcal{L}]_{j,j}[\mathcal{L}^{n}]_{j,1}\nonumber \\
&+\sum_{\ell \in S^j_c} [\mathcal{L}]_{j,\ell}[\mathcal{L}^{n}]_{\ell,1},
\end{align}
where $q$ is the parent of $j$, and $S^j_c$ is the set of the children of node $j$. Since $d(q,1) \geq n+1$, $d(j,1) \geq n+2$, and $d(\ell,1) \geq n+3$ for $\ell \in S^j_c$, due to the induction assumption we conclude $[\mathcal{L}^{n}]_{q,1}=0$, $[\mathcal{L}^{n}]_{j,1}=0$, and $[\mathcal{L}^{n}]_{\ell,1}=0$, respectively. Therefore, $[\mathcal{L}^{n+1}]_{j,1}=0$.
\item We next consider node $m$ where $d(m,1)=n+1$. Using (\ref{eq_im}) we can write
\begin{align}
[\mathcal{L}^{n+1}]_{m,1}=&[\mathcal{L}]_{m,q'}[\mathcal{L}^{n}]_{q',1}+[\mathcal{L}]_{m,m}[\mathcal{L}^{n}]_{m,1}\nonumber \\
&+\sum_{\ell \in S^m_c} [\mathcal{L}]_{m,\ell}[\mathcal{L}^{n}]_{\ell,1},
\end{align}%
where $q'$ is the parent of $m$, and $S^m_c$ is the set of the children of $m$. Since $[\mathcal{L}^n]_{m,1}=[\mathcal{L}^n]_{\ell,1}=0$ for $\ell \in S_c^m$, and $[\mathcal{L}^n]_{q',1}=\mathcal{W}(P_{q',1})$ from the induction assumption, it can be concluded that $[\mathcal{L}^{n+1}]_{m,1}=w_{m,q'}\mathcal{W}(P_{q',1})=\mathcal{W}(P_{m,1})$. 
\end{itemize}
\hfill $\blacksquare$
 
\begin{figure}[!bthp]
\centering
\includegraphics[scale=0.250]{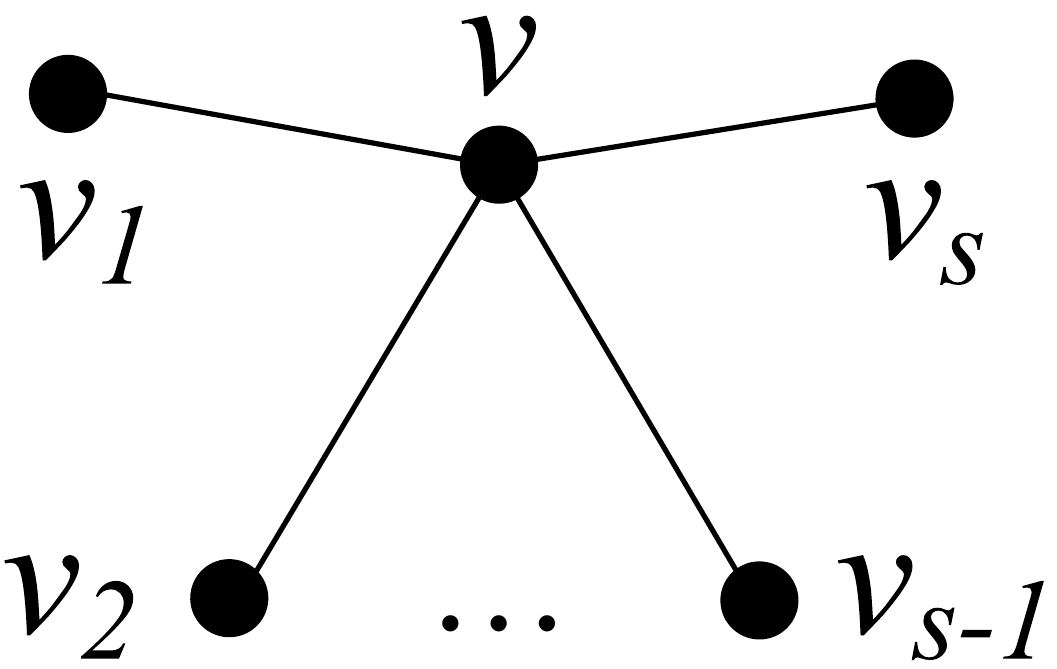}
\caption{The subgraph $\mathcal{H}$ defined in Lemma \ref{lemma5}}
\label{fig_lemma5}
\end{figure}

\begin{lemma}
\label{lemma5}
Consider a node indexed as $v$ in a weighted graph $\mathcal{G}$ and its neighboring nodes denoted by $v_1$ through $v_s$. Let $\mathcal{H}$ be a subgraph of $\mathcal{G}$ induced by the set of all edges incident to $v$ as illustrated in Fig. \ref{fig_lemma5}. Let $\mathcal{L}=-L$, where $L$ is the weighted Laplacian matrix of $\mathcal{G}$. \textcolor{black}{Let $\mathcal{W}(\mathcal{H})$ denote the weights of all edges belonging to $\mathcal{H}$. Then,    $[\mathcal{L}^i]_{v_s,1}$ can be uniquely computed from $\mathcal{W}(\mathcal{H})$ and $[\mathcal{L}^i]_{m,1}$, ($m \in \mathcal{V}(\mathcal{H})\backslash \{ v_s \}$), $\forall~i \geq 1$. }
\end{lemma}
\proof Noting that (\ref{eq_im}) holds for $\mathcal{L}$ belonging to any type of graphs (cyclic and acyclic), we can write: 

\vspace{-1em}
\begin{footnotesize}
\begin{align*}
[\mathcal{L}^{i+1}]_{v,1}&=[\mathcal{L}]_{v,v}[\mathcal{L}^{i}]_{v,1}+ [\mathcal{L}]_{v,v_1}[\mathcal{L}^{i}]_{v_1,1}+\cdots+[\mathcal{L}]_{v,v_s}[\mathcal{L}^{i}]_{v_s,1}.
\end{align*}%
\end{footnotesize}%
Therefore,
\begin{align}
[\mathcal{L}^{i}]_{v_s,1}&=\frac{N_1}{[\mathcal{L}]_{v,v_s}},
\label{eq_thir2}
\end{align}   
where $N_1=[\mathcal{L}^{i+1}]_{v,1}-[\mathcal{L}]_{v,v}[\mathcal{L}^{i}]_{v,1}-([\mathcal{L}]_{v,v_1}[\mathcal{L}^{i}]_{v_1,1}+\cdots+[\mathcal{L}]_{v,v_{s-1}}[\mathcal{L}^{i}]_{v_{s-1},1})$. \textcolor{black}{The RHS of (\ref{eq_thir2}) is a function of $[\mathcal{L}^i]_{m,1}$, ($m \in \mathcal{V}(\mathcal{H})\backslash \{ v_s \}$) and $\mathcal{W}(\mathcal{H})$. }\hfill$\blacksquare $
 
{\bf Note:} For node $v_s$, $[\mathcal{L}^{i}]_{v_s,1}$ $\forall~i \geq 1$ is uniquely identifiable if $[\mathcal{L}^i]_{m,1}$, ($m \in \mathcal{V}(\mathcal{H})\backslash \{ v_s \}$) and $\mathcal{W}(\mathcal{H})$ is identifiable. Such nodes, from this point onward, will be referred to as {\it available} nodes. This definition of \emph{availability} will be used as a critical argument for the forthcoming proofs.   
%%%%%%%%%%%%%%%%%%%%%%%%%%%%%%%%%%%%%%%%%%%%%%%%%%%%%%%%

Using Lemmas \ref{lemma2} and \ref{lemma5}, we next propose a simple hierarchical algorithm to design the output matrix $C$ in (\ref{eq1}) that guarantees global identifiability of $W$.

%%%%%%%%%%%%%%%%%%%%%%%%%%%%%%%%%%%%%%%%%%%%%%%%%%%%%%%%
%Using the above lemmas, We next propose a simple hierarchical algorithm for placing sensors in $\mathcal{T}$ as follows:
\begin{algorithm}[!ht] \caption{\label{al1} Sensor Placement Algorithm for Acyclic Consensus Networks}
\begin{algorithmic}[1]
%	\Require  Measurements $y_{q_j}(t_k),\; t_k = t_0,\; t_1, \ldots, t_{\varsigma}$ and nominal keys $\mathcal{K}_{j} = [\mathcal{K}_{1j}, \ldots,\mathcal{K}_{pj}] $
\State Partition $\mathcal{V}(\mathcal{T})$ into sets of $S_0$ to $S_p$.
\State Start with $S_0$ and place a sensor at this node.
\For {$k=1 \to p$}
\For {each set of siblings $S^j_k$} 
\State choose any $|S^j_k|-1$ nodes belonging to $S^j_k$ and 
\\ \hspace{2.7em} place sensors at them.  %Accordingly, if $|S^k_j|=1$ for some $k$ then no sensor is needed for that single node.
\EndFor 
\EndFor  
\end{algorithmic}
\end{algorithm}

\begin{example}
The sensor placement steps of Algorithm \ref{al1} are illustrated through an example shown in Fig. \ref{fig_2}. The different steps of the placement are shown in Figs. \ref{fig2_1} to \ref{fig2_4}. Circles around nodes indicate that a sensor is placed there. Step 1 puts a sensor at node 1 (input node). In step 2, we do not need to put any sensor at node 2 because $|S^1_1|=1$. In step 3, we put one sensor in either node 3 or node 5 since $|S^1_2|=2$. Say, we choose node 3. In the final step we put a sensor in either node 4 or node 6 since $|S^1_3|=2$. Say, we choose node 6. 
Our claim is that the state measurements from these three sensors are sufficient to guarantee unique identifiability of the edge weights of $\mathcal{T}_1$. We next state Theorem \ref{theorem1} to justify this claim for any rooted tree graph $\mathcal{T}$.
\begin{figure}[!thb]
\centering
\subfloat[$\mathcal{T}_1$]{\includegraphics[scale=0.40]{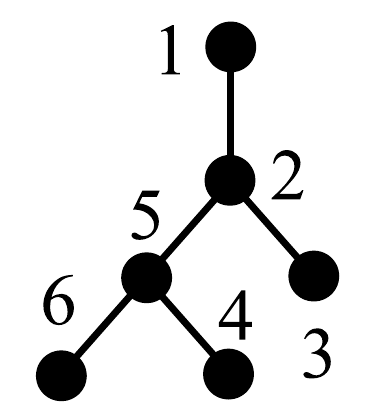}\label{fig_1b}}~
\subfloat[Step 1]{\includegraphics[scale=0.40]{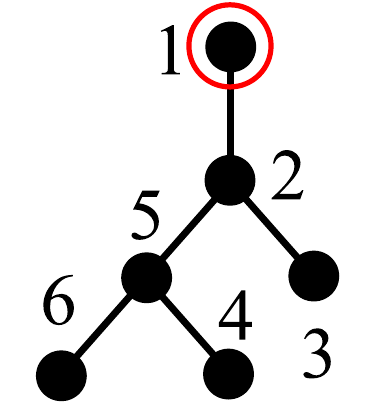} \label{fig2_1}}~
\subfloat[Step 2]{\includegraphics[scale=0.40]{p1_0.pdf}\label{fig2_2}}~
\subfloat[Step 3]{\includegraphics[scale=0.40]{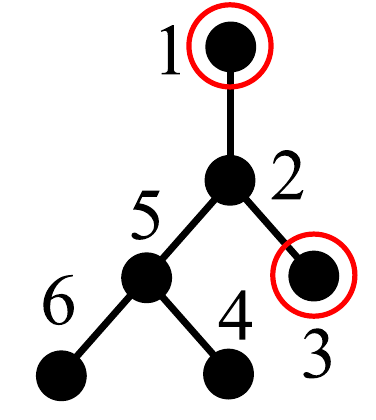}}~
\subfloat[Step 4]{\includegraphics[scale=0.40]{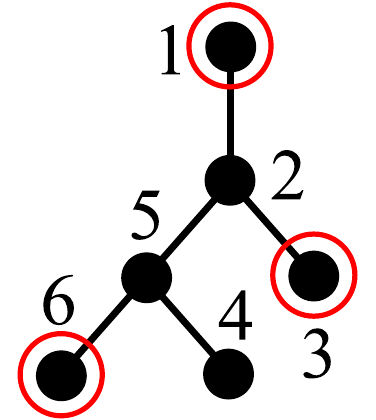}\label{fig2_4}}
\caption{Application of Algorithm \ref{al1} to $\mathcal{T}_1$ (circles indicate sensor nodes)}
\label{fig_2}
\end{figure}
\end{example}

\begin{theorem}
\label{theorem1}
Consider the consensus model (\ref{eq1}) with $\mathcal{G}=\mathcal{T}$. Let $\mathcal{S} \subset \mathcal{V}(\mathcal{T})$ be a set of sensor nodes determined by Algorithm \ref{al1}, $\mathbf{y}(t)$ be the corresponding output measured by $\mathcal{S}$, and $H(W)$ be the transfer function from $u(t)$ to $\mathbf{y}(t)$. Then, the mapping from the $W$ to $H(W)$  is one-to-one.
\end{theorem}
 \proof
Let us partition $W$ into sets $W_{0,1}$, $W_{1,2}$ through $W_{p-1,p}$ where%\footnote{\textcolor{black}{We again emphasize that in the entire proof, $w_{i,j}=w_{j,i}$ for all $i,j$. }} 
\begin{align}
W_{j,j+1}&=\{w_{u,v} \in W;~ u\in S_j,~v\in S_{j+1}\}.
\label{eq_W}
\end{align}
Let $Q_0$ through $Q_{p}$ denote the Markov parameters of (\ref{eq1}) defined as
\begin{align}
Q_j\triangleq C\mathcal{L}^jB,~~~j=0,\ldots,p.
\label{markov}
\end{align}
We show by strong induction that the mapping from $W_{j,j+1}$ to $\bigcup_{j=0}^{2n-1}Q_{j}$ is one-to-one if sensors are placed in $\mathcal{T}$ following Algorithm \ref{al1}. 
\begin{figure}[!tbhp]
\centering 
\includegraphics[scale=0.85]{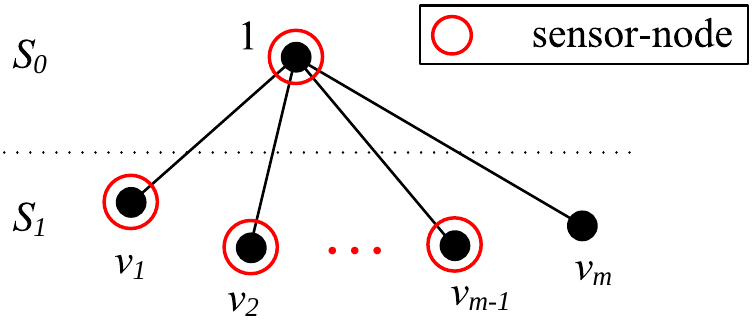}
\caption{$S_0 \cup S_1$}
\label{S1}
\end{figure}
\newline {\bf Step 1:} We first show that the mapping from $W_{0,1}$ to $Q_1$ is one-to-one. For this, let us consider $\tilde{Q}_1 \subset Q_1$, where  
$\tilde{Q}_1\triangleq \mathbf{e}^T_k\mathcal{L}\mathbf{e}_1=[\mathcal{L}]_{k,1}$, and $k$ is the index of the measured nodes in $S_0 \cup S_1$. For example, assume that the  nodes in $S_1$ are indexed as $v_1, v_2, \ldots, v_{m}$, where $m=|S_1|$ as shown in Fig. \ref{S1}. Then $k=1,v_1,\ldots,v_{m-1}$, and 
\begin{align}
\tilde{Q}_1=\begin{bmatrix} \left [ \mathcal{L} \right ]_{1,1} \\ \left [ \mathcal{L} \right ]_{v_1,1}\\ \vdots \\ \left [ \mathcal{L} \right ]_{v_{m-1},1} \end{bmatrix}=\begin{bmatrix}
-(w_{1,v_1}+\cdots+w_{1,v_{m}})\\ w_{1,v_1} \\ \vdots \\ w_{1,v_{m-1}}
\end{bmatrix}.
\label{eq_7}
\end{align}
\noindent From (\ref{eq_7}), it can be easily seen that the mapping from $W_{0,1}$ to $\tilde{Q}_1 \subseteq Q_1$ is one-to-one. Also, based on Lemma \ref{lemma5} and considering the subgraph induced by $W_{0,1}$, the term  $[\mathcal{L}^i]_{v_{m},1}$ $\forall~i\geq 1$ is uniquely identifiable. In other words, $v_m$ is an \emph{available} node. \hfill $\Box$

\begin{figure}[!htpb]
\centering
\includegraphics[scale=0.75]{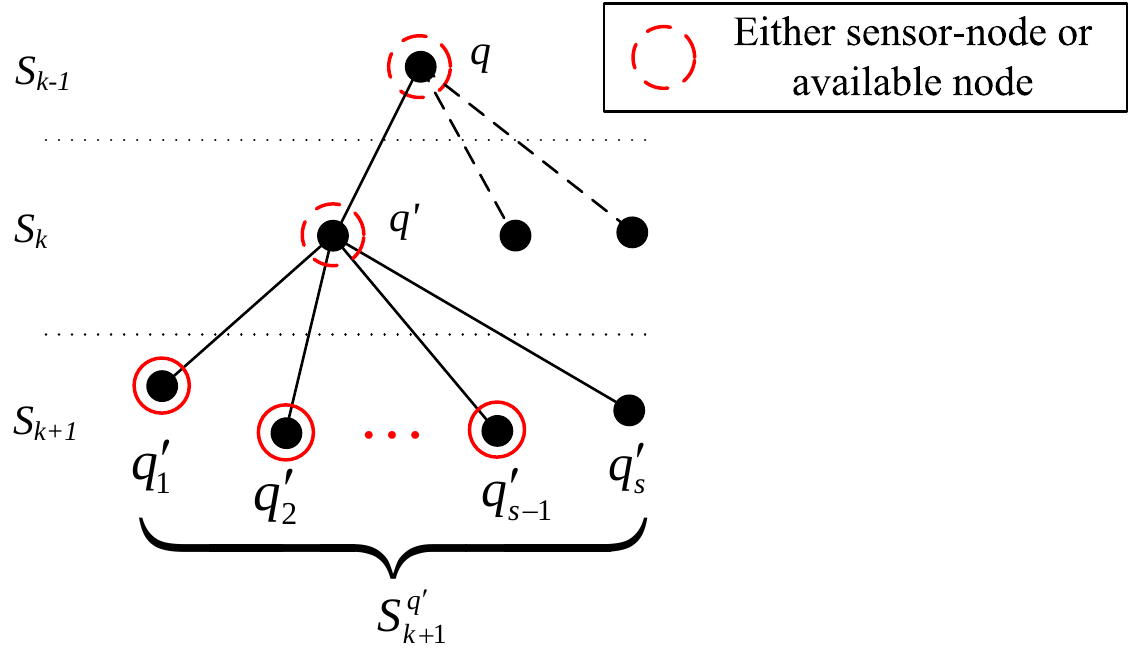}
\caption{$S_{k+1}^{q'}$, the set of children of node $q' \in S_k$}
\label{fig_3}
\end{figure}

\noindent {\bf Step 2:} Let us assume by strong induction that $W_{i-1,i}$ for $i \leq k$ (for some $k>1$), and all non-sensor nodes $v \in S_0\cup \ldots \cup S_{k}$ are \emph{available} nodes. Note that, this also implies that $\mathcal{W}(P_{1,i})$ for all nodes $i \in S_0 \cup \ldots \cup S_{k}$ is identifiable. We next prove that $W_{k,k+1}$ will be uniquely identifiable from $\bigcup_{j=0}^{2n-1}Q_{j}$. Let us consider the sibling set $S^{q'}_{k+1}$ as shown in Fig. \ref{fig_3}. The elements of this sibling group are indexed as $q'_1,\ldots,q'_s$.    

\noindent {\bf Step 2.1:} First, the identifiablity of $w_{q',q'_1}$ through $w_{q',q'_{s-1}}$ will be proved. Let us consider $\tilde{Q}_{k+1} \subset Q_{k+1}$ as $\tilde{Q}_{k+1}$=$\mathbf{e}^T_{\ell} \mathcal{L}^{k+1}\mathbf{e}_1$ for $\ell=q'_1,\ldots,q'_{s-1}$. We can write 

\begin{small}
\begin{align}
\tilde{Q}_{k+1}&=\begin{bmatrix}
\left [ \mathcal{L}^{k+1} \right ]_{q'_1,1}\\ \vdots \\ \left [ \mathcal{L}^{k+1} \right ]_{q'_{s-1},1}
\end{bmatrix} \nonumber \\
&\overset{\textrm{Lemma \ref{lemma2}}}{=}\begin{bmatrix}
\mathcal{W}(P_{1,q'_1})\\ \vdots \\ \mathcal{W}(P_{1,q'_{s-1}})
\end{bmatrix}=\mathcal{W}(P_{1,q'})\begin{bmatrix}
w_{q',q'_1}\\ \vdots \\w_{q',q'_{s-1}}
\end{bmatrix}.
\label{eq_72}
\end{align}
\end{small}%
Since $\mathcal{W}(P_{1,q'})\neq 0$ is uniquely identifiable from the induction assumption, from (\ref{eq_72}) we conclude that $w_{q',q'_1}$ through $w_{q',q'_{s-1}}$ are uniquely identifiable. 

\noindent {\bf Step 2.2:} We next prove the identifiability of $w_{q',q'_s}$. This will be done by considering the term $[\mathcal{L}^{k+1}]_{q',1}$. If node $q'$ has a sensor, then $[\mathcal{L}^{k+1}]_{q',1}$ is a subset of $Q_{k+1}$. If it does not have a sensor, then based on the induction assumption  it is an \emph{available} node, and $[\mathcal{L}^{k+1}]_{q',1}$ is uniquely identifiable from $\bigcup_{j=0}^{2n-1}Q_{j}$. We recall (\ref{eq_im}) as
\begin{align}
[\mathcal{L}^{k+1}]_{q',1}=&[\mathcal{L}]_{q',q}[\mathcal{L}^{k}]_{q,1}+[\mathcal{L}]_{q',q'}[\mathcal{L}^{k}]_{q',1}\nonumber \\
&+\sum_{\ell \in S^{q'}_{k+1}} [\mathcal{L}]_{q',\ell}[\mathcal{L}^{k}]_{\ell,1}.
\label{14}
\end{align}
Based on Lemma \ref{lemma2}, $[\mathcal{L}^{k}]_{\ell,1}=0$ for $\ell \in S^{q'}_{k+1}$. Thus, considering $[\mathcal{L}]_{q',q'}=-\sum_{\ell \in \mathcal{N}_{q'}} w_{q',\ell}$, we can rewrite (\ref{14}) as

\vspace{-1em}
\begin{footnotesize}
\begin{align}
w_{q',q'_s}=-w_{q',q}-\sum_{\ell=1}^{s-1}w_{q',q'_\ell}-\frac{[\mathcal{L}^{k+1}]_{q',1}-[\mathcal{L}]_{q',q}[\mathcal{L}^{k}]_{q,1}}{[\mathcal{L}^{k}]_{q',1}}, \label{eq_13}
\end{align}
\end{footnotesize}%
where $[\mathcal{L}^{k}]_{q',1} =\mathcal{W}(P_{q',1})\neq 0$ from Lemma \ref{lemma2}. The terms $[\mathcal{L}^{k}]_{q',1}$ and $[\mathcal{L}^{k+1}]_{q',1}$ are either the Markov parameters of the system, if a sensor is placed at node $q'$, or identifiable from the Markov parameters by the induction assumption ($q'$ is available). The same argument is applicable to $[\mathcal{L}^{k}]_{q,1}$. The term $[\mathcal{L}]_{q',q} = w_{q',q}$ and other edge weights in the RHS of (\ref{eq_13}) are also identifiable from the induction assumption and Step 2.1. Thus, $w_{q',q'_s}$ is identifiable.

\noindent {\bf Step 2.3:} The final step is to show that node $q'_s$ is an \emph{available} node. This follows directly from Lemma \ref{lemma5} by considering the subgraph induced by the edges incident to~$q'$.

\textcolor{black}{{\bf Remark:} 
It should be noted that, as stated in Algorithm \ref{al1}, it does not matter which $|S_{k+1}^{q'}|-1$ nodes will be chosen out of $|S_{k+1}^{q'}|$ nodes in each step.  This happens due to the fact that, based on Lemma \ref{lemma5}, any non-sensor node becomes an available node after placing sensors in all its siblings. 
}

Steps 2.1, 2.2, 2.3 can be generalized to all sets of siblings belonging to $S_{k+1}$, which concludes the proof for the induction. Since the mapping from the edge weights to Markov parameters are shown to be one-to-one, then the mapping from the edge weights to the transfer function from $u(t)$ to $\mathbf{y}(t)$ is also one-to-one. This statement is equivalent to saying that the edge-weights are identifiable from the input-output data.  
%\begin{table}[H]
%\centering
%\caption{Mapping Structure from Edge Weight Sets to Markov Parameters}
%\label{tab_tri}
%\begin{tabular}{|c|c|c|c|c|c|c|}
%\hline
%Parameter & $Q_1$ & $Q_2$ & $Q_3$ & $Q_4$ & $\cdots $ & $Q_{p+1}$ \\
%\hline $W_{0,1}$ & $\surd$   &    & & & &\\
%\hline $W_{1,2}$ & $\surd$   & $\surd$  &$ \surd$ & & & \\
%\hline $W_{2,3}$ & $\surd$   & $\surd$  & $\surd$ &$\surd$ & & \\
%\hline $\vdots$ & &  & & & $\ddots$ & \\
%\hline $W_{p-1,p}$ & $\surd$   & $\surd$ & $\surd$ & $\surd$ & $\cdots$ & $\surd$\\
%\hline
%\end{tabular} 
%\end{table} 
\hfill $\blacksquare$

\textcolor{black}{Theorem \ref{theorem1} ensures that the edge-weights $W$ are identifiable from the input-output data provided that sensors are placed using Algorithm 1. In particular, once this is satisfied any identification algorithm, for example least squares, may be used to identify $W$. Moreover, it is not necessary to estimate the Markov parameters and then estimate the $W$ from them.}  %The entire proof is only to show the identifiability and not the identification step.}

%We conclude this section by stating the number of sensors needed for Algorithm \ref{al1}. 
\subsection{Number of sensors needed}
\textcolor{black}{ Recall that a leaf of a tree $\mathcal{T}$ is a vertex that has only one adjacent node. Let $L_\mathcal{T}$ denote the \emph{non-input leaves} of $\mathcal{T}$, i.e., the set of leaves that are not the input node.} %Using these definitions, we next state the number of sensors needed by Algorithm 1.}
\begin{prop}
\label{pro2} If Algorithm \ref{al1} is applied to a rooted-tree $\mathcal{T}$ then the number of placed sensors is equal to $L_\mathcal{T}$.  
\end{prop}
\proof The proof follows from the induction on $r$, the number of nodes of $\mathcal{T}$.

Step 1 ($r=2$): In this case $L_\mathcal{T}=1$. Also, from Algorithm \ref{al1} it is clear that only one sensor is needed, that being at the input node. 

Step 2 ($r=k$ where $k > 2$ is an integer): Let us assume that the number of required sensors is equal to $L_\mathcal{T}$, and prove the same is true when one more node is added to $\mathcal{T}$. For this, let us form the tree $\mathcal{T}'$ by adding a new vertex $v'$ and its incident edge $e'$ to $\mathcal{T}$. Two further cases can arise:

1. If $v'$ is a neighbor of $v$, a non-input leaf of $\mathcal{T}$, then $v$ will not be a non-input leaf of $\mathcal{T}'$ anymore but $v'$ will be a new non-input leaf, and therefore  $L_{\mathcal{T}'}=L_\mathcal{T}$. In this case, no extra sensor is required to be added to satisfy identifiability of $W_{\mathcal{T}'}$ based on Algorithm \ref{al1}.
 
2. If $v'$ is a neighbor of $\bar{v}$ (an internal vertex of $\mathcal{T}$ or the input node), then $v'$ will be a new non-input leaf of $\mathcal{T}'$ and $L_{\mathcal{T}'}=L_\mathcal{T}+1$. Also, for this case Algorithm \ref{al1} stipulates addition of a new sensor at $v'$. 

Steps 1 and 2 verify that Algorithm \ref{al1} results in a sensor placement scheme with $L_\mathcal{T}$ number of sensors. \hfill $\blacksquare$ \\
\subsection{Example}    
In the example of Fig. \ref{fig_2}, $\mathcal{T}$ has three non-input leaves, namely, nodes 3, 4, and 6. The number of sensors needed by Algorithm \ref{al1} is also 3. However, the choice of the sensors is not unique. For example, any of the sets $S_1=\{ 1,3,4 \}$, $S_2=\{1,3,6\}$, $S_3=\{1,5,4\}$, and $S_4=\{1,5,6\}$ will guarantee identifiability of the edge sets for $\mathcal{T}$ of Fig. \ref{fig_2}. Further examples will be shown in Section \ref{sec_ex}.

\subsection{More Information About the Edge-Weights from a Transfer Function}
Let us assume a sensor is placed at any arbitrary node $i \in \mathcal{V}(\mathcal{T})$, $y(t)$ be the corresponding measured output, and $H(s,W)$ be the transfer function from $u(t)$ to $y(t)$. The question is what combinations (or functions) of the edge-weights will be identifiable. 
%We next show that the mapping from the $\sum_{w \in W(\mathcal{T})} w$ and the mapping from $\mathcal{W}(P_{i,1})$ to $H(s,W)$ are one-to-one, where $\mathcal{W}(P_{i,1})$ is the weight of the unique path of length $d(i,1)$ connecting nodes $i$ and $1$. 
Assuming that (\ref{eq1}) is controllable and observable with $C=\mathbf{e}^T_i$ and $B=\mathbf{e}_1$, then $H(s,W)=C\big(sI-\mathcal{L}(W)\big)^{-1}B$ can be rewritten as

\vspace{-1em}
\begin{small}
\begin{align}
H(s,W)=\frac{b_1s^{n-1}+b_2s^{n-2}+\cdots+b_n}{s^n+a_1s^{n-1}+\cdots+a_n}.
\end{align} %
\end{small}%
From \cite[Theorem 2.1]{fad}, $a_1=-\mathrm{trace}(\mathcal{L})=2(\sum_{w \in W(\mathcal{T})} w)$. Thus, the mapping from the $\sum_{w \in W(\mathcal{T})} w$ to $H(s,W)$ is one-to-one. Also, recalling the results of Lemma \ref{lemma2}, $[\mathcal{L}^k]_{i,1}=\mathcal{W}(P_{i,1})$ if $k=d(i,1)$. Since $[\mathcal{L}^k]_{i,1}$ is a Markov parameter of the system corresponding to $y(t)$, the mapping from $\mathcal{W}(P_{i,1})$ to $H(s,W)$ is one-to-one. Thus, we are able to find at least two functions of $W$ that are identifiable from $H(W)$. However, investigating the identifiability of the individual edge weights may not be possible from a single sensor for any general network. %The reason is finding the exact injective mappings of the edge weights to the remaining coefficients of $H(s,W)$ become intractable in general.
\begin{table*}[!hbpt]
\caption{Examples of sensor placement for acyclic graphs based on Algorithm \ref{al1}. $N$ denotes the number of sensors. The listed Markov parameters are chosen to show their one-to-one relationship with the unknown edge weights. }
\label{tab1}
\centering
\begin{tabular}{|c|c|l|l}
\hline \noindent\parbox[c][1em][c]{8 em}{\centering Network Example} & \noindent\parbox[t][1em][t]{0.5 em}{ \centering $N$} & \noindent\parbox[b][1em][c]{47 em}{\centering Markov Parameters Needed for Identifiability Proof} \\
\hline \includegraphics[scale=0.65]{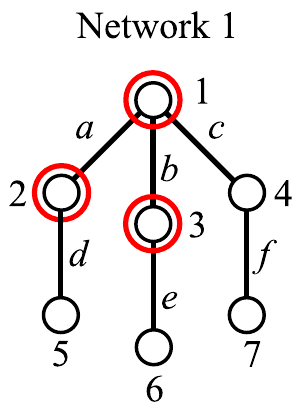} & \noindent\parbox[b][10em][c]{0.5 em}{ \centering $3$} & \noindent\parbox[b][10em][c]{17 em}{\begin{footnotesize}  $\begin{matrix}[l]
[\mathcal{L}]_{2,1}~~=f_1(a)=a\\ 
[\mathcal{L}]_{3,1}~~=f_2(b)=b\\
[\mathcal{L}]_{1,1}~~=f_3(a,b,c)=-(a+b+c)\\
[\mathcal{L}^2]_{2,1}=f_4(a,b,c,d)=- a(2a+ b + c) - ad \\
[\mathcal{L}^2]_{3,1}=f_5(a,b,c,e)=- ab - bc - be - 2b^2\\
[\mathcal{L}^3]_{1,1}=f_6(a,b,c,d,e,f)=- 4a^3 - 5a^2b  - 5a^2c  - da^2 - 5ab^2 - 6abc - 5ac^2   - 4b^3 - 5b^2c - eb^2\\
\hspace{5.0em} - 5bc^2 - 4c^3 - fc^2
\end{matrix}$ 
\end{footnotesize} }  \\
\hline \hspace{-0.1em} \includegraphics[scale=0.65]{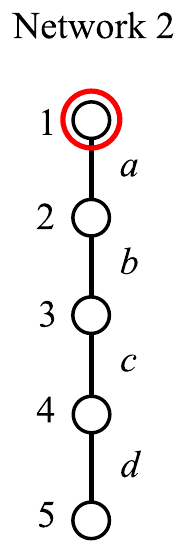}& \noindent\parbox[b][11em][c]{.5 em}{ \centering $1$} & 
\noindent\parbox[b][11em][c]{17 em}{\begin{footnotesize} $\begin{matrix}[l]
\mathrm{[} \mathcal{L} \mathrm{]}_{1,1}~~=f_1(a)=a\\
\mathrm{[} \mathcal{L}^3 \mathrm{]}_{1,1}=f_2(a,b)=- 4 a^3 - b a^2\\
\mathrm{[} \mathcal{L}^5 \mathrm{]}_{1,1}=f_3(a,b,c)=- 16 a^5 - 12 a^4 b - 9 a^3 b^2 - 4 a^2 b^3 - c a^2 b^2\\
\mathrm{[} \mathcal{L}^7 \mathrm{]}_{1,1}=f_4(a,b,c,d)=- 64 a^7 - 80 a^6 b - 73 a^4 b^3 - 12 a^4 b^2 c - 44 a^3 b^4  - 8 a^3 b^2 c^2 - 16 a^2 b^5 - 12 a^2 b^4 c  \\
\hspace{5.0em}   - 4 a^2 b^2 c^3 - d a^2 b^2 c^2 - 88 a^5 b^2 - 18 a^3 b^3 c - 9 a^2 b^3 c^2
\end{matrix} $ \end{footnotesize}}\\
\hline \includegraphics[scale=0.62]{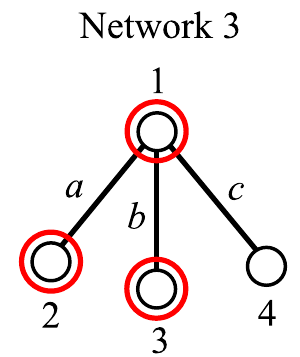}& \noindent\parbox[b][7em][c]{.5 em}{ \centering $3$} & \noindent\parbox[b][7em][c]{17 em}{$\begin{matrix}[l]
[\mathcal{L}]_{2,1}~=a\\
[\mathcal{L}]_{3,1}~=b\\
[\mathcal{L}]_{1,1}~=-a-b-c
\end{matrix}$} \\
\hline 
\end{tabular}
\end{table*}
 
%%%%%%%%%%%%%%%%%%%%%%%%%%%%%%%%%%%%%%%%%%%%%%%%%
%%%%%%%%%%%%%%%%%%%%%%%%%%%%%%%%%%%%%%%%%%%%%%%%%   
\section{Examples}
\label{sec_ex}
Table \ref{tab1} shows  three examples to illustrate how our proposed sensor placement algorithm provide one-to-one mapping from the edge weights to the Markov parameters. We also show that depending on the graph topology, our algorithm may become necessary and not just sufficient for identifiability.
\subsection{Illustrating the one-to-one mapping between weights and the Markov parameters}
Let us consider three acyclic networks shown in Table \ref{tab1}. The first compartment of Table \ref{tab1} shows the network. The second compartment shows $N$, the number of the sensors that are placed in each network. The third compartment lists the respective Markov parameters following from Algorithm \ref{al1} that are needed to prove the identifiability for each example. The sequence in which the Markov parameters are listed is important. From all of these lists, we can see that every line item introduces exactly one unknown. This means that the Markov parameters may be nonlinear functions of the edge weights, but the mapping between the two is one-to-one. 

\subsection{Sufficiency vs Necessity}
\label{sec_ex2}
We next show that depending on the topology of a tree graph, Algorithm \ref{al1} may be either sufficient, or both necessary and sufficient for identifiability. For example, we consider two acyclic graphs, Networks 2 and 3 cited in Table \ref{tab1}. Based on Algorithm \ref{al1}, Network 2 only requires a single sensor to be identifiable, which is clearly the minimum possible number of sensors. Therefore, Algorithm \ref{al1} in this case is clearly both necessary and sufficient. For Network 3, however, Algorithm \ref{al1} is only sufficient but is not necessary. We can check this by removing the sensor placed at node 1. It can be shown that $a$, $b$, and $c$ are still identifiable in this situation. For example, consider the following three Markov parameters, $[\mathcal{L}]_{2,1}=a$, $[\mathcal{L}]_{3,1}=b$, and $[\mathcal{L}^2]_{2,1}=- 2a^2 - ab - ac$ that show that the mapping from the parameters $(a,b,c)$ to the Markov parameters is one-to-one. Clearly, Algorithm \ref{al1} in this case places one extra sensor than necessary. However, that does not mean that one can remove any one arbitrary sensor from Network 3 of Table \ref{tab1}, and still preserve the identifiability property. For example, if a careless user removes the sensor from node 3, the network will still have the minimum number of the sensors needed to be identifiable, but will not be identifiable anymore. The unidentifiability of the edge-weights, in this case, can be shown by considering another set of weights for network 3, i.e.,  $w_{1,2}=a$, $w_{1,3}=c,$ $w_{1,4}=b$ that produces the same output and Markov parameters as Network 3 with edge-weights shown in Table \ref{tab1}. This small example nicely illustrates that the location of the sensor nodes in identifiability is equally important as the number of sensors.       
%\input{Eqn1/e48}
%It can be easily verified that the mapping from $(a,b,c)$ to the coefficients of $H(s)$ is one-to-one. This is an example to show that Algorithm \ref{al1} provides a sufficient number of sensors, which is not always necessary. 
\textcolor{black}{In case $a=b=c$, then it can be easily verified that network 3 is uncontrollable. However, the minimal subsystem still contains sufficient information for identifying ($a=b=c$) uniquely.} %mapping form the edge-weights of this network to the coefficients of the minimal input-output transfer functions is still one-to-one.% These minimal transfer functions are as follows:
%\begin{subequations}
%\begin{align}
%H_1(s)&=\frac{X_1(s)}{U(s)}=\frac{s+c}{s^2+s(a+b+2c)},\\
%H_2(s)&=\frac{X_2(s)}{U(s)}=\frac{a}{s^2+s(a+b+2c)},\\
%H_3(s)&=\frac{X_3(s)}{U(s)}=\frac{b}{s^2+s(a+b+2c)}.
%\end{align}
%\end{subequations}
%It can be easily verified that the edge-weights are identifiable from the minimal transfer functions of $H_1(s)$, $H_2(s)$, and $H_3(s)$ despite the uncontrollabality of the system.  }

\section{A Note on Star Graphs}
\label{sec_star}
\begin{figure}[!h]
\centering
\includegraphics[scale=0.43]{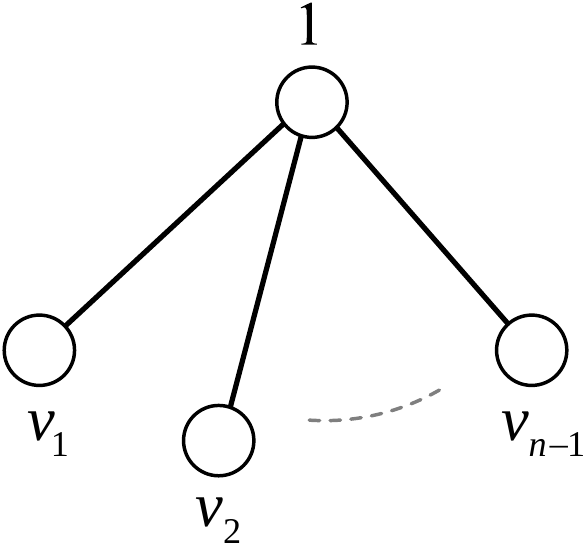}
\vspace{-1em}
\caption{The star graph of Proposition \ref{prop_star}}
\label{fig_star}
\vspace{-1em}
\end{figure}

\textcolor{black}{Let us consider the star graph with $n$ nodes shown in Fig. \ref{fig_star}. It can be easily verified that Algorithm \ref{al1} assigns $(n-1)$ sensors to guarantee identifiability. We next show that although this number looks conservative, Algorithm \ref{al1} actually puts only one extra sensor than the number of sensors that is necessary identify this network.  }
\begin{prop}
\label{prop_star}
If the system (\ref{eq1}) is defined over the star graph $\mathcal{T}$ of Fig. \ref{fig_star}, then the minimum number of sensors to identify this network is $(n-2)$. %Let the input node to be the central node of the star\footnote{the node that is adjacent to all other nodes of the star graph} and be indexed as 1. If $n-2$ sensors are placed in all nodes of the graph excluding node 1 and any other nodes of the graph, then all edges of the network are identifiable. If we remove any other node from this network, then the network will become unidentifiable. 
\end{prop}
\proof
We first prove that with $(n-3)$ sensors, the edge weights of $\mathcal{T}$ are not identifiable. The $(n-3)$ sensors can be placed in either of the following cases: 

Case 1. Suppose sensors are placed in all nodes other than three nodes chosen from the set $\{v_1,v_2,\ldots,v_{n-1}\}$. Say we choose the nodes $v_2,v_3,v_4$. In this case the following edge-weights are not identifiable: $w_{1,v_1}$, $w_{1,v_2}$, and $w_{1,v_3}$. To show unidentifiability we can easily show that the following two edge-weight parameters:

\vspace{-1em}
\begin{small}
\begin{align*}
w_{1,v_1}=a,&~ w_{2,v_2}=b,~ w_{1,v_3}=c\\
w_{1,v_1}=a,&~ w_{2,v_2}=c,~ w_{1,v_3}=b, 
\end{align*}
\end{small}%
are not distinguishable. 

Case 2. Suppose sensors are placed in all nodes other than node 1 and two nodes chosen from the set $\{v_1,v_2,\ldots,v_{n-1}\}$. Say we choose the nodes $v_1,v_2$. In this case the following edge-weights are not identifiable: $w_{1,v_1}$, $w_{1,v_2}$ as shown by the inditinguishability of the following two edge-weight parameters:

\vspace{-1em}
\begin{small}
\begin{align*}
w_{1,v_1}=a,&~ w_{2,v_2}=b,\\
w_{1,v_1}=b,&~ w_{2,v_2}=a.
\end{align*}
\end{small}%
So far we have proved $(n-3)$ sensors are not sufficient for identifiability of the edge-weights of $\mathcal{T}$, we next prove that $(n-2)$ sensors are sufficient for identifiability of the edge weights of $\mathcal{T}$. Assume $(n-2)$ sensors are placed at all nodes of $\mathcal{T}$ excluding node 1 (the input node) and any other nodes of the graph (say node $v_{n-1}$), then it can be shown that all edges of the network are identifiable. For this, it suffices to consider the following set of Markov parameters:

\vspace{-1em}
\begin{small}
\begin{align*}
&[\mathcal{L}]_{v_1,1}=w_{v_1,1},\\
&~~~~~~~~~\vdots\\
&[\mathcal{L}]_{v_{n-2},1}=w_{v_{n-2},1},\\
&[\mathcal{L}^2]_{v_1,1}=-w_{v_1,1}(w_{v_1,1}+\cdots+w_{v_{n-1},1})-w^2_{v_1,1},
\end{align*}
\end{small}%
which proves the one-to-one mapping from the edge weights to this Markov parameter set.  \hfill $\blacksquare$

\section{Conclusions}
\label{con}
In this paper, we developed a sensor placement algorithm to ensure global identifiability of weighted consensus networks with first-order dynamics and tree structures. We showed that the proposed algorithm provide a sufficient condition for identifiability of the edge weights of acyclic network graphs by proving a one-to-one mapping from the edge weights to the Markov parameters of the system. The method, however, becomes intractable for any generic cyclic graph. We also derive the number of the sensors needed for identifiability, and show that depending on the graph topology, this number may be more than necessary for certain graphs. Our algorithm provide simple yet sufficient ways of placing sensors in large consensus networks for accurate, real-time identification. Our future direction of research is to generalize these findings for networks with arbitrary cyclic structures and for consensus networks with differential-algebraic models.

%%%%%%%%%%%%%%%%%%%%%%%%%%%%%%%%%%%%%%%%%%%%%%%%%%

\end{document}